\documentstyle[prc,aps,preprint]{revtex}

\draft
\begin{document}

\title{Numerical Convergence in Solving the Vlasov Equation}
\author{C.\ Jarzynski and G.F.\ Bertsch}
\address{Institute for Nuclear Theory, University of Washington,
             Seattle, WA 98195, USA}
\date{\today}
\maketitle

\begin{abstract}
When the Vlasov equation is investigated numerically
using the method of test particles,
the particle-particle interactions that inevitably arise
in the simulation (but are not present in
the Vlasov equation itself) result in an accumulation
of errors which eventually drive the
collection of test particles toward a state of
classical thermal equilibrium.
We estimate the rate at which these errors accumulate.
\end{abstract}

\vspace{.5in}

The Vlasov equation plays a central role in
classical (and semiclassical) time-dependent mean field
theory, and has been used to model a wide variety of
many-body processes, from the gravitational $N$-body
problem\cite{grav}, to plasma physics\cite{plasma},
to nuclear dynamics\cite{nuclear}.
While the content of the Vlasov equation is conceptually
simple --- interactions among many
particles are replaced by a common mean-field potential ---
solutions are harder to come by, and must in general
be sought numerically.
This is often accomplished with the test particle
method: a swarm of numerical particles
is used to simulate a distribution $f({\bf r},{\bf p},t)$
in one-body phase space, and the mean-field potential
in which these test particles evolve is obtained from
this distribution.
Thus, while the Vlasov equation replaces a physical
many-body problem with the self-consistent evolution
of a one-body phase-space distribution, the test
particle method in turn replaces the Vlasov equation
with a numerical many-body problem.
This raises the issue of convergence:
for a given number of test particles, and over
a given length of time, how closely can we expect the
evolution of $f({\bf r},{\bf p},t)$ as obtained by
the test particle method, to resemble the true evolution
under the Vlasov equation?
In the limit of arbitarily long evolution time, we can
certainly expect the test particle method to fail.
In that limit, the unavoidable interactions between
individual test particles will drive the swarm of
numerical particles toward a Boltzmann distribution
of energies,
whereas under the Vlasov equation there are no
particle-particle interactions, and $f$ typically does
not evolve toward classical thermal equilibrium.

A relevant example of this disagreement
arises in the application of the Vlasov equation to
nuclear dynamics, where the Pauli principle is
imposed by insisting that, initially,
$f\le 4/h^3$ everywhere in phase space.
Under the Vlasov equation, this condition is preserved
exactly with time; with the test particle method,
however, classical thermalization occurs, and the
Pauli condition is violated.
(See Fig.\ 4 of Ref.\cite{rs}, I, for an illustration.)

The eventual thermalization of test particles may
occur on a time scale longer than that in which
one is interested.
Nevertheless, it is indicative of a general process,
whereby interactions between the test particles
introduce errors which drive the numerical
solution away from the actual solution of the Vlasov
equation.
It it thus important to obtain an estimate of the
rate at which these errors accumulate.
Such an estimate is the goal of the present brief
note.

We will restrict ourselves mainly to self-consistent
potentials which are local functions of particle
density, although a brief discussion of long-range
forces will be presented at the end.
Working with a simple schematic model, we will
argue that errors in the particle energies
accumulate diffusively, and we will solve for the
functional dependence of the associated diffusion
constant $D_E$, in terms of physical and numerical
parameters.
For the specific case where Gaussian smoothing is used
to obtain the particle density $\rho$ from the
positions of the test particles, we obtain a more
quantitative prediction for $D_E$.
Finally, we compare our theoretical predictions
with numerical results.

The Vlasov equation is explicitly given by
\begin{equation}
{\partial f\over\partial t}\,+\,\{f,H\}\,=\,0\,,
\end{equation}
where $\{\cdot,\cdot\}$ denotes the ordinary Poisson
bracket, and the Hamiltonian $H$ is
\begin{equation}
H({\bf r},{\bf p})\,=\,{p^2\over 2m}\,+\,
U_f({\bf r}).
\end{equation}
The notation $U_f({\bf r})$ is meant to indicate that
the mean-field potential $U({\bf r})$ is a functional
of the one-body phase space distribution $f$.
Often (e.g.\ when the physical interactions between
particles are independent of momentum) the functional
dependence of $U$ on $f$ reduces to a dependence only on
the density $\rho$ in ordinary space:
\begin{eqnarray}
U_f({\bf r})\,&\rightarrow&\,
U_\rho({\bf r})\\
\rho({\bf r})\,&=&\,\int d{\bf p}\,
f({\bf r},{\bf p}).
\end{eqnarray}
Throughout this paper, we will assume, for simplicity,
that this is the case.
Note that if $U$ is linear in $\rho$, then it may
be expressed in terms of a two-body physical
interaction $V_{12}$:
$U_\rho({\bf r})=\int d{\bf r}^\prime\,
\rho({\bf r}^\prime)\,V_{12}({\bf r},{\bf r}^\prime)$.
In general, however, $U$ need not be linear in $\rho$,
i.e.\ the mean field need not arise from two-body
interactions.

The implementation of the test particle method involves
two tasks:
(1) evolving each of the $N$ test particles in the
presence of the time-dependent potential $U$; and
(2) constructing $U$ from the positions of the
particles at any instant in time.
The first is straightforward, involving simply the
numerical integration of Hamilton's equations of
motion.
The second task requires the particle density
$\rho({\bf r},t)$, which is obtained by smearing
the position of each point particle with a
localized folding function $g$:
\begin{equation}
\label{eq:density}
\rho({\bf r},t)\,=\,{A\over N}\,\sum_{i=1}^N
g({\bf r}-{\bf r}_i(t)).
\end{equation}
Here $A$ is the number of physical particles,
whereas the sum runs over the test particles,
located at positions ${\bf r}_i(t)$ at time $t$.
$g({\bf r})$ is a function localized in a volume
$\sigma^3$ around the origin,
and normalized to unity: $\int d{\bf r}\,g({\bf r})=1$.
The parameter $\sigma$ thus measures the distance
over which we smear out the particle positions.
Gaussian folding functions are commonly used.

To estimate the rate of accumulation of errors
introduced by interactions among the test particles,
let us consider a simple model in which our many-particle
system is confined within a box of volume $V$.
Furthermore, let us take the functional dependence
of $U$ on $\rho$ to be local:
\begin{equation}
\label{eq:local}
U_\rho({\bf r})=U(\rho({\bf r})).
\end{equation}
That is, the potential at {\bf r} depends only on
the density of particles at that point;
this corresponds to zero-range interactions
among particles, and is commonly used to model
short-range interactions such as nuclear forces
(see e.g.\ the Skyrme parametrization\cite{nuclear}).
It is important to distinguish here between the
mathematical problem one is trying to solve
(propagation under the Vlasov equation), and the
numerical method used to solve it:
even if the potential $U_\rho({\bf r})$
which enters into the Vlasov equation is exactly
local --- as indeed we are assuming --- the
interactions between test particles in a numerical
implementation will necessarily have finite range,
due to the smearing which is employed to extract a
smooth density $\rho({\bf r})$ from the positions
of a finite number of test particles.

Now, consider an initial phase space distribution
$f_0({\bf r},{\bf p})$ corresponding to an ensemble
of monoenergetic particles distributed uniformly
throughout the box, with an isotropic distribution
of momenta.
Explicitly, this has the form
$f_0({\bf r},{\bf p})\propto\delta(p-p_0)\Theta_B({\bf r})$,
where $p\equiv\vert{\bf p}\vert$,
and $\Theta_B({\bf r})$ is equal to 1 (0) if {\bf r}
is inside (outside) the box.
As can be seen by inspection, this phase space
distribution is a stationary solution of the
Vlasov equation,
thus under the Vlasov equation the ensemble
of particles remains exactly monoenergetic.
Our strategy now will be to investigate how such
an initial distribution evolves under a numerical
simulation using test particles.
Specifically, after a time $\Delta t$, what is the
amount $\Delta E$ by which the energy of a typical
test particle has strayed from its initial value?
The growth of $\Delta E$ with $\Delta t$ will then
be a measure of the accumulation of error inherent
in the test particle method.

Let us take our $N$ test particles --- all given the
same initial speed $v=p_0/m$ --- to be distributed
randomly throughout the container, and choose
$\sigma$ so that $V/N\ll\sigma^3\ll V$.
This will result in a reasonably smooth numerical
density $\rho({\bf r},t)$,
without smearing over too large a volume of the box.
We can express this density as
\begin{equation}
\rho({\bf r},t)\,=\,\rho_0\,+\,
\delta\rho({\bf r},t),
\end{equation}
where $\rho_0=A/V$ is the physical density which we are
trying to simulate, and $\delta\rho({\bf r},t)$ represents
the fluctuations around $\rho_0$ due to the finite number
of test particles.
To gauge the typical size of $\delta\rho$, note that
the value of $\rho$ at a given point is roughly
equal to $(A/N)\, n/\sigma^3$, where $n$ is the number of test
particles within a volume $\sigma^3$ of the point in
question.
(The factor $A/N$ is a conversion factor between the density
of test particles and the density of physical particles.)
On average, $n$ will be given by $n_0=N\sigma^3/V$,
with fluctuations of size $\sqrt{n_0}$ around this average.
These considerations yield the following expression
for the typical size of the fluctuations $\delta\rho$:
\begin{equation}
\label{eq:rhorms}
\delta\rho_{rms}\,\sim\,{A\over N}{\sqrt{n_0}\over\sigma^3}
\,=\,{\rho_0\over\sqrt{n_0}}.
\end{equation}
It should be clear as well that
$\delta\rho({\bf r}_1,t)$ and $\delta\rho({\bf r}_2,t)$
will be correlated only if ${\bf r}_1$ and ${\bf r}_2$
are within a distance $\sim\sigma$ of one another.
Furthermore, at a given location {\bf r}, the value of
$\delta\rho({\bf r},t)$ will be correlated over a time
$t_c\sim\sigma/v$, since that is a typical time over
which a test particle remains within a volume
element $\sigma^3$ of {\bf r}.

Thus, our numerical density $\rho({\bf r},t)$
fluctuates in space and time around an
average value $\rho_0=A/V$, where the size of
the fluctuations is given by
$\delta\rho_{rms}\sim\rho_0/\sqrt{n_0}$,
and these fluctuations are correlated over
a distance $\sigma$, and a time $t_c\sim\sigma/v$.
Let us now make use of this picture to determine
what happens to a given test particle evolving
under the potential $U(\rho({\bf r},t))$
computed from this numerical density.

We first expand the potential $U$ around its value
at $\rho_0$:
\begin{equation}
U(\rho({\bf r},t))\,=\,
U(\rho_0+\delta\rho({\bf r},t))\,\approx\,
U_0\,+\,U^\prime_0\,\delta\rho({\bf r},t),
\end{equation}
where $U_0\equiv U(\rho_0)$ and
$U_0^\prime\equiv {dU\over d\rho}(\rho_0)$.
Thus, like the numerical density $\rho$,
the potential $U$ fluctuates in space and time
around an average value ($U_0$).
The typical rate at which $U$ is changing, at
a fixed point {\bf r}, is determined by the typical
rate of change of $\delta\rho$:
\begin{equation}
\label{eq:dudt}
\Biggl\vert{\partial\over\partial t} U({\bf r},t)
\Biggr\vert\,\sim\,
U_0^\prime\,{\delta\rho_{rms}\over t_c}.
\end{equation}

Now consider a single test particle $i$ moving
under this time-dependent potential.
{}From Hamilton's equations,
the rate of change of the total energy $E_i$ of the
particle is exactly the value of $\partial U/\partial t$
along its trajectory:
\begin{equation}
\dot E_i(t)\,=\,
{\partial U\over\partial t}({\bf r}_i(t),t).
\end{equation}
This function $\dot E_i(t)$ is autocorrelated
over a time scale $t_c\sim\sigma/v$,
which is considerably shorter than a characteristic
time scale associated with the particle's motion
in the box (e.g.\ the traversal time across the
length of the box).
The change in energy $\Delta E_i$ is thus the
time integral of a function $\dot E_i(t)$ which
fluctuates rapidly, with short time correlations;
this implies that $\Delta E_i$ evolves diffusively.
The associated diffusion constant $D_E$ is then
the time integral of the auto-correlation
function of $\dot E_i(t)$.
Approximating this integral by the product of the
mean-square value of $\dot E_i(t)$ with the correlation
time $t_c$, we have, using Eq.\ref{eq:dudt},
\begin{equation}
D_E\,\sim\,\Bigl\vert\dot E_i\Bigr\vert^2\,{\sigma\over v}
\,\sim\,
\Bigl(U_0^\prime\,\delta\rho_{rms}\Bigr)^2\,
{v\over\sigma}.
\end{equation}
Finally, using $\delta\rho_{rms}\sim\rho_0/\sqrt{n_0}$,
and $n_0=N\sigma^3/V$, we get
\begin{equation}
\label{eq:diffcon}
D_E\,\sim\,(U_0^\prime)^2\rho_0 v\,
\cdot{A\over N}\cdot {1\over\sigma^4}.
\end{equation}
Thus after a time $\Delta t\gg t_c$, we can expect
the energy of our test particle to have changed
by an amount
\begin{equation}
\label{eq:deltae}
\Delta E\,\sim\,(D_E\,\Delta t)^{1/2},
\end{equation}
with $D_E$ given by Eq.\ref{eq:diffcon} above.
Eqs.\ref{eq:diffcon} and \ref{eq:deltae} together
describe the accumulation of error in the
energy of a typical test particle, and thus
constitute our main result.

Note that we have written $D_E$ as the product of
three factors, the first of which contains
only physical quantities, while the other two depend
on purely numerical parameters: the smearing parameter
$\sigma$, and the number of test particles per
physical particle, $N/A$.
The prediction that the error accumulates more
slowly for larger values of $N/A$ is expected;
this is the benefit of using more test particles.
Eq.\ref{eq:diffcon} predicts that one
gains even more by increasing the value of the smearing
parameter $\sigma$: $D_E\propto\sigma^{-4}$.
As pointed out by Reinhard and Suraud\cite{rs} (II),
this should not come as a surprise: a larger smearing
effectively suppresses the interaction between individual
test particles, thus slowing the rate at which energy
gets exchanged.
Of course, smearing distorts the mean field itself;
therefore too much of it, while suppressing errors due
to test particle interactions, will result in an inaccurate
simulation of the Vlasov equation.
Ultimately, one wants $\sigma$ large enough so that
$\Delta E$ remains small over the time scale of
physical interest, but not so large as to distort the
inhomogeneities that are physically present in the
mean field.

As mentioned earlier, one expects that in the long
run the test particles thermalize.
This ought to happen on a time scale $\tau$ over which
each test particle has had the opportunity to change
its energy by an amount comparable to the average
particle energy, $mv^2/2$.
Thus,
\begin{equation}
\label{eq:thermcon}
(D_E\,\tau)^{1/2}\,\sim\, mv^2.
\end{equation}
Combining this with Eq.\ref{eq:diffcon},
we obtain for the thermalization time scale
\begin{equation}
\label{eq:main}
\tau\,\sim\,
{m^2v^3\over (U_0^\prime)^2\rho_0}\,\cdot\,
\sigma^4\,\cdot\,{N\over A}\quad.
\end{equation}
In numerical experiments aimed at studying the
relaxation toward thermal equilibrium under the
test particle method, Reinhard and Suraud have
found that doubling the value of $\sigma$
``gains more than an order of magnitude in the
relaxation time'' (Ref.\cite{rs}, p.\ 227);
this is in agreement with our prediction here
that $\tau$ scales like $\sigma^4$.
Furthermore, these authors have predicted that
$\tau\propto N/A$, and have confirmed this
numerically.

A few comments are now in order.
First, in a realistic test particle simulation, the
particles are held together by the mean field itself,
rather than being artificially confined within a box.
Nevertheless,
the mechanism by which the test particles exchange
energy with one another remains the same, therefore the
result derived within the context of our simple model
ought to hold in the more realistic situation as well.

Next, while our main result predicts how the growth
of errors scales with the various parameters involved,
a more quantitative estimate will depend on the details
of how the test particles interact with one another.
For instance, the use of a gaussian folding function,
$g({\bf r})=(2\pi\sigma^2)^{-3/2}\exp (-r^2/2\sigma^2)$,
allows for an explicit evaluation of $\delta\rho_{rms}$.
This leads to an expression for $D_E$ which
has the form of Eq.\ref{eq:diffcon}, but with
a numerical factor $1/8\pi^{3/2}$ in front.
Alternatively, for gaussian folding functions one can
evaluate (within the linear approximation) the amount
of energy exchanged in a given collision between two
test particles, in terms of impact parameter.
The further assumption that different particle-particle
collisions are uncorrelated leads (after some work)
to a diffusion coefficient
\begin{equation}
\label{eq:quant}
D_E\,=\,{1\over 12\pi}\,
(U_0^\prime)^2\rho_0 v\,
\cdot{A\over N}\cdot {1\over\sigma^4}.
\end{equation}
It is encouraging that this approach, which differs
somewhat from that leading to Eq.\ref{eq:diffcon},
nevertheless yields the same functional dependence
of $D_E$ on the various quantities involved.

Finally, our assumption that the dependence of $U$
on $\rho$ is local (Eq.\ref{eq:local}) was made
both for the sake of simplicity, and because our
original motivation to study this problem arose from the
application of the Vlasov equation to nuclear dynamics,
where short-range physical forces lead to a local $U_\rho$.
However, in many physical applications of the Vlasov
equation one deals with long-range forces
(e.g.\ Coulombic and gravitational forces), therefore
it may be useful to extend the analysis of the present
work, to include non-local mean-field potentials.
It is interesting in this context to note that
Chandrasekhar\cite{chandra} has made a detailed
calculation of the time scale $T_E$ required for binary
stellar interactions to drive a self-gravitating
system of many stars (e.g.\ a galaxy) toward thermal
equilibrium.
His result, translated into our notation, is
$T_E\sim m^2 v^3/(Gm^2)^2\rho$,
where we have removed dimensionless factors.\footnote{
including the logarithm of a quantity which is essentially
the ratio of kinetic to potential energy for a typical
star in the system.}
Now, the gravitational potential at a given point
in the galaxy is roughly $U\sim -NGm^2/R$, where $R$ is
a distance scale characterizing the size of the galaxy,
and $N$ is the number of stars.
If, for purposes of comparison with Eq.\ref{eq:main} above,
we write $U^\prime\sim U/\rho\sim U/NR^{-3}$, then
we get
\begin{equation}
T_E\,\sim\,
{m^2 v^3\over(U^\prime)^2\rho}\,\cdot\,R^4.
\end{equation}
This has the form of our Eq.\ref{eq:main}, only
with $\sigma$ replaced by $R$, and without the
factor $N/A$ (since there are no ``test particles'').
The fact that the size of the entire galaxy,
$R$, appears in place of our smoothing parameter
$\sigma$, suggests that --- in comparison with
short-range forces --- long-range forces such
as gravity strongly suppress the collisional
relaxation toward thermal equilibrium.
This is consistent with numerical findings:
in semiclassical simulations of nuclear
dynamics, the (undesirable) approach to
classical Boltzmann equilibrium often takes place
on a time scale comparable with the mean-field
dynamics in which one is interested\cite{epla}.
In contrast, simulations of the many-body
gravitational problem evolve rapidly to a
near-static ``collisionless equilibrium'',
which differs from a microcanonical
distribution\cite{kandrup}.

We now present the results of numerical experiments
which we have performed to test our predictions.
We simulated the schematic model discussed
above --- a gas of particles confined to a
box --- where the box was taken to be a cube of
volume 1000 (in arbitrary units) with periodic
boundary conditions, and the mean field potential
was taken as $U(\rho)=-.6\rho+4\rho^2$.
$A=200$ physical particles of mass $m=1$ were assumed,
the initial speed of each particle was set to $v=1$,
and a gaussian folding function was used.
In each simulation, we allowed the gas of particles
to evolve for a time $\Delta t=2.5$, and we followed
the growth of the mean square change in test particle
energy,
$\langle(\Delta E)^2\rangle\equiv
(1/N)\sum_{i=1}^N(\Delta E_i)^2$,
over this time.
For each simulation, we found $\langle(\Delta E)^2\rangle$
to grow linearly with time\footnote{aside from the
inevitable short quadratic growth at the start}, in
agreement with Eq.\ref{eq:deltae}.
To extract a numerical energy diffusion coefficient
$D_E$, we divided the final value of
$\langle(\Delta E)^2\rangle$ by $\Delta t$.

We ran two sets of simulations.
In the first set, the smearing parameter was
held fixed at $\sigma = 1$, and the number of
test particles was varied from $N=2000$ to
$N=10000$.
In Fig.\ref{fig:na} we plot the resulting values
of $D_E$ as a function of $N/A$ on a log-log plot.
{}From Eq.\ref{eq:diffcon}, we expect the data to
fall along a straight line of slope -1;
the solid line gives the best fit of the data
to a line of this slope.
In the second set of simulations, the number of
test particles was held
fixed at 4000, and the smearing parameter was
varied from $\sigma=.5$ to $\sigma=1$.
Fig.\ref{fig:sigma} plots the resulting values
of $D_E$ as a function of $\sigma$ on a log-log plot,
and the solid line gives the best fit of the data
to a line of slope -4.
In both figures we find good agreement between the
prediction of Eq.\ref{eq:diffcon} and the numerical
results.
If we allow the slope of the lines to vary as well,
then a best fit of straight lines to the two sets of
data yields slopes of $-1.08$ and $-4.54$, instead
of $-1$ and $-4$.
Thus, the discrepancy between the predicted slope
and the best fit to numerical data is on the order
of $10\%$ in each case.

Since we used a gaussian folding function in our
numerical simulations, it is interesting to
compare the results directly to the quantitative
prediction of Eq.\ref{eq:quant}.
That prediction is depicted by the dotted line
in each of the figures.
We see that Eq.\ref{eq:quant} overestimates the
value of $D_E$ by a factor of nearly 2.
We believe that this discrepancy is due to our
neglect of correlations between different collisions.
(Since motion in our mean field is highly regular,
we can expect that the energy exchanged in a
given collision between two test particles will
be somewhat correlated with the energy exchanged
at the next collision of those two particles.)

To conclude, we have argued in this paper that
when the Vlasov equation is studied numerically
using the method of test particles, the inevitable
interactions between test particles introduce
errors which accumulated diffusively.
Our main result --- a prediction of how the associated
diffusion constant scales with the available numerical
parameters --- was shown to agree well with the
results of computer simulations.

We acknowledge conversations with P.G.\ Reinhard and
J.\ Randrup which stimulated and improved this work.
This work was supported by the Department of Energy
under Grant No.\ DE-FG06-90ER40561.

\begin{figure}
\caption{The energy diffusion constant $D_E$ as a function
of $N/A$, the number of test particles per physical particle.
The heavy dots show the values of $D_E$ extracted from numerical
experiments, the solid line shows the best fit of these
points to a straight line of slope -1,
and the dotted line shows the quantitative prediction
of Eq.\ref{eq:quant}.}
\label{fig:na}
\end{figure}

\begin{figure}
\caption{$D_E$ as a function of smearing parameter $\sigma$.
The solid line is a best fit of the data to a straight line
of slope -4;
the dotted line represents Eq.\ref{eq:quant}.}
\label{fig:sigma}
\end{figure}

\end{document}